# Evolution of Electronic and Magnetic Properties in Topological Semimetal $SmSb_xTe_{2-x}$

Krishna Pandey[1], Rabindra Basnet[2], Jian Wang[3], Bo Da[4], Jin Hu[1,2*]

[1]Materials Science and Engineering Program, Institute for Nanoscience and Engineering, University of Arkansas, Fayetteville, Arkansas 72701, USA

[2]Department of Physics, University of Arkansas, Fayetteville, Arkansas 72701, USA

[3]Department of Chemistry, Wichita State University, Wichita, Kansas 67260, USA

[4]Research and Services Division of Materials Data and Integrated System, National Institute for Materials Science, 1-1 Namiki, Tsukuba, Ibaraki 305-0044, JAPAN

## Abstract

The ZrSiS-type materials have attracted intensive attention due to the existence of various topological fermions. The magnetic version of the ZrSiS-type materials, *Ln*SbTe (*Ln* = lanthanides), is an ideal candidate to explore novel exotic states due to the interaction between magnetism and topology. In this work, we report the experimental study on structural, magnetic, thermodynamic, and electronic properties for $SmSb_xTe_{2-x}$ with various Sb content. The revealed evolutions of these properties with tuning the compositions would provide useful insights for the fundamental topological physics and the future applications.

[*]Email: jinhu@uark.edu

## 1. Introduction

The discoveries of topological semimetals offer unprecedented opportunities to explore entirely new classes of materials and develop wide range of next generation device applications. The electronic states of these materials possess symmetry protected, linearly dispersed Dirac or Weyl crossings, hosting electrons whose low energy excitations can be described by Dirac or Weyl equations [1–4]. Among various topological semimetals, the topological nodal line semimetal (NLSM) exhibits interesting linear band crossing along a one-dimensional loop or line near the Fermi level [5]. Belonging to NLSMs, the ZrSiS-family compounds show rich phenomena due to the presence of two types of Dirac states: the gapless Dirac point state protected by non-symmorphic symmetry, and the slightly gapped Dirac nodal-line state generated by glide-mirror symmetry [6–13]. Such a large material family can be represented by a chemical formula *WHM* (*W*= Zr/Hf, *H*= Si/Ge/Sn/Sb, *M* = S/Se/Te) [6,9,13–18], which crystallizes in a PbFCl-type crystal structure with space group *P*4/*nmm*. The various combinations of *W*, *H* and *M* provide great tunability for spin-orbit coupling and structural dimensionality. Exotic properties, such as the unusual surface floating state [19,20], electronic correlation enhancement [21–23], and pressure-induced topological phase transitions [24,25] has been discovered.

In addition to non-magnetic *WHM* compounds, the magnetic version *Ln*SbTe (*Ln* = lanthanides) exhibits long range magnetic order brought in by magnetic *Ln* [26–33], and thus provides a platform to study the interplay between magnetism and topological states [33]. In addition to magnetism, rich quantum phenomena such as Kondo effect, charge density waves (CDWs), and correlation enhancement have been reported in various *Ln*SbTe (*Ln*= Ce, Nd, Sm, Gd, Ho) compounds [26,27,29–31,33–36]. More specifically, the AFM ground state has been observed in all reported *Ln*SbTe compounds except the non-magnetic LaSbTe [26–28,30,31,33,34,36–40]. Despite of similarities in structure and the existence of off-stoichiometric compositions, different properties have been found in various magnetic *Ln*SbTe. For example, non-metallic transport with Kondo like features has been reported in CeSbTe [27,34], NdSbTe [31] and SmSbTe [36], which is distinct from the reported metallic transport in LaSbTe [37] and GdSbTe [26,28], HoSbTe [30]. In addition, large Sommerfeld coefficient have been found in specific heat measurements on NdSbTe [31], HoSbTe [30] and SmSbTe [36], which is very different from the low Sommerfeld coefficient in CeSbTe [27,33] and GdSbTe [28]. The topological Dirac states has been established in stoichiometric *Ln*SbTe compounds [32–34,36,39,41], which also been probed in off-stoichiometric GdSbTe compounds [26]. Furthermore, the Dirac nodal line in $GdSb_xTe_{2-x-\delta}$ has been found to be robust despite of the changes in composition and structure [29,39]. The interplay of crystal symmetry, magnetism, band topology and electron correlations is expected to drive into various topological states [26,33,36,41,42]. Further, crystal structure, magnetic phases, and electronic states are found to be tunable with the composition stoichiometry in $LnSb_xTe_{2-x}$ [26,29,42]. The distortion in Sb-square net cause the modification in charge density waves, electronic band structure and inherent magnetism [26,29,42]. Thus, the off-stoichiometry in $LnSb_{2-x}Te_x$ provide a novel route to study the magnetic phases, topological states and their interplay can provide a platform which can induce different topological states at different value of *x* [26,29,33].

With this motivation, in this work we present a comprehensive study of the evolution of structural, electronic, and magnetic properties with varying composition in $SmSb_xTe_{2-x}$ ($0 < x \leq 1$). The stoichiometric or nearly stoichiometric SmSbTe has been recently identified to be a magnetic topological semimetal [36,41] exhibiting a combination of a few interesting properties such as Dirac nodal-line fermions, enhanced electronic correlations, antiferromagnetic ground states with possible magnetic frustration, and Kondo effects [36]. In this work, by varying the composition stoichiometry we have revealed the tetragonal to orthorhombic structural phase transition and the evolution of electronic and magnetic properties in $SmSb_xTe_{2-x}$. With such tunable properties, $SmSb_xTe_{2-x}$ provides a good platform to study and design various quantum states.

## 2. Experiment

The single crystals of $SmSb_xTe_{2-x}$ ($0 < x \leq 1$) were synthesized by a two-step chemical vapor transport method similar to that for growing the stoichiometric SmSbTe crystals [36]. To obtain crystals with different Sb content *x*, the ratio of Sb and Te in source materials was varied. Millimeter-size single crystals with metallic luster can be obtained using this method, as shown in the inset of Fig. 1(a). The compositions of each $SmSb_xTe_{2-x}$ single crystal sample used in this work were carefully determined by energy dispersive x-ray spectroscopy (EDS) scans on multiple spots. The crystal structures were determined by single crystal x-ray diffraction (SCXRD). For each sample in this study, we used the EDS composition because XRD cannot precisely distinguish Sb and Te due to their similar electronic configurations. The magnetic, thermal, and electronic properties of $SmSb_xTe_{2-x}$ were measured by using a physical properties measurement system (PPMS).

## 3. Results and Discussions

The composition analysis using EDS indicates that the Sb/Te ratios of the obtained crystals are usually less than that in source materials, which has also been widely observed in this family of materials [35,43]. The excellent crystallinity of our single crystals is demonstrated by the sharp (00*l*) x-ray diffraction peaks, as shown in Fig. 1(a). The systematic low-angle shift of these (00*l*) peaks upon increasing Sb content indicates elongated *c*-axis. The complete structural information of $SmSb_xTe_{2-x}$ was further determined by structural refinement using single crystal XRD, as shown in Table I. Our structural analysis reveals a structural transition with varying Sb content *x*. As shown in Fig. 1b, upon increasing the Sb content near $x \sim 0.8$, $SmSb_xTe_{2-x}$ undergoes a structure transition from the orthorhombic space group *Pmmm* (no.47) (green color) to the tetragonal *P*4/*nmm* (no. 123) (yellow color), which is accompanied by an elongation of *c*-axis and a shrinkage of *ab*-plane. The structure change is more visualized by a parameter of *c*/*a*. For Sb-less composition compound, $SmSb_{0.11}Te_{1.85}$, *c*/*a* is 2.099. The Sb-more composition compound, $SmSb_{0.93}Te_{1.07}$, the *c*/*a* is 2.155. With increasing Sb content *x*, the *c*/*a* follows an obvious increasing trend. Similar structure phase transition has also been reported in other $LnSb_xTe_{2-x}$ (*Ln*= La, Ce, Gd) compounds [29,42,43].

We depict the crystal structures for tetragonal phase in Fig. 1c and illustrated the structural differences with orthorhombic phases in Fig. 1d. As shown in Fig. 1c, the crystal structure of $SmSb_xTe_{2-x}$ is characterized by a layered structure, with each layer consisting of a Sb plane sandwiched by Sm-Te bilayers. In nonstoichiometric $SmSb_xTe_{2-x}$ samples, the Sb plane is partially substituted by Te. The orthorhombic distortion in Sb-less ($x < 0.8$) samples is manifested by the distort Sb layer. As shown in Fig. 1d, the tetragonal sample ($SmSb_{0.93}Te_{1.07}$) features a 2D Sb square-net with identical Sb-Sb bonding length and 90° bonding angles. For the orthorhombically distorted lattice, though the Sb-Sb bonding length remains the same within each sample, the bonding angles deviate from the 90°. With reducing Sb content, such deviation in bonding angles becomes stronger, which is in line with the greater difference between lattice parameters *a* and *b* (Fig. 1b). Furthermore, the Sb atoms are not located in the same 2D plane since the same-side interior angles are not supplementary. This can also be seen from the atomic positions summarized in Table I. In addition to the orthorhombic distortion induced by reducing Sb content, previous studies have also revealed the existence of vacancies in the Sb layer in other *Ln*SbTe compounds, which is reported to cause atomic orders and charge density waves (CDWs) [29,34,35]. Similar Sb vacancies have also been found in our materials by SCXRD refinement in which the occupancy of the Sb layer was released and refined. The refinement also reveals a suppression of the amount of vacancies with increasing the Sb content. For example, the occupancy of the partially substituted Sb layer is found to be 93(±1)% for the Sb-less $SmSb_{0.11}Te_{1.85}$ sample, while 99(±1)% for the Sb-rich $SmSb_{0.93}Te_{1.07}$ sample. Such trend is also consistent with our composition analysis by EDS, and has been reported in the related compound GdSbTe [29]. Though our SCXRD refinement indeed reveals the existence of vacancies, it should be noted that the solid clarification of vacancies in $SmSb_xTe_{2-x}$ samples is difficult due to the instrument limitation of our *x*-ray diffractometer.

The evolution of magnetism with Sb content was studied using magnetization and heat capacity measurements. Some *Ln*SbTe compounds (*Ln* = Ce, Gd, Ho) and their off-stoichiometric forms $LnSb_xTe_{2-x}$ have been found to show antiferromagnetic ground states with metamagnetic transitions [27,29–31,33,34,39,42]. In $SmSb_xTe_{2-x}$, the temperature dependent magnetic susceptibility $\chi(T)$ measured under both in-plane (*H*//*ab*) and out-of-plane (*H*//*c*) magnetic field orientations shows peaks at low temperatures (Fig. 2a). The absence of irreversibility (data not shown) between zero-field cooling (ZFC) and field cooling (FC) indicates an AFM nature of the magnetic order, which is further supported by the negative Curie-Weiss temperature as will be shown below. Our previous study has revealed a Néel temperature $T_N \approx 3.7$ K for tetragonal, stoichiometric SmSbTe, which does not change with the applied magnetic field [36]. Similar field-independent $T_N$ has also been observed in both orthorhombic (Fig.2b) and tetragonal (Fig.2c) off-stoichiometric compounds, which is distinct from many other *Ln*SbTe compounds such as CeSbTe [27,33,42] NdSbTe [31,40], GdSbTe [29,35] and HoSbTe [30].

In non-stoichiometric $SmSb_xTe_{2-x}$ studied in this work, interestingly, multiple magnetic phase transitions have been observed for samples with the Sb content $x \geqslant 0.36$ and up to the tetragonal/orthorhombic phase boundary of $x = 0.8$. Data for a typical example ($SmSb_{0.62}Te_{1.37}$) is shown in Fig. 2b, in which two transition-like features indicated by $T_{N1}$ and $T_{N2}$ can be observed. The two transitions can be better resolved in the derivative of susceptibility $d\chi(T)/dT$ (Fig. 2b, inset). As will be shown later, the emergence of two peaks is also observed in heat capacity measurements, in which $T_{N1}$ corresponds to a low temperature broad peak while $T_{N2}$ corresponds to a shaper peak at high temperatures. Multiple ordering temperatures has also been observed in many structurally similar rare earth compounds such as $CeTe_2$, $SmTe_2$, $GdTe_2$, as well as $GdSb_xTe_{2-x-\delta}$ with higher Sb deficiency [29,44], which is possibly associate with the various CDW vectors or vacancy distributions in orthorhombic phases [42,45]. In Figs. 2d we summarized the composition dependence of the magnetic ordering temperature for $SmSb_xTe_{2-x}$. Overall, $T_{N1}$ does not change strongly with composition variation, while $T_{N2}$ grows with increasing Sb content $x$ and suddenly disappears in the tetragonal phases with $x > 0.8$.

We have extracted Curie-Weiss temperature ($\theta_{cw}$) and effective magnetic moments ($\mu_{eff}$) by fitting the data in the paramagnetic phase using a modified Curie-Weiss model $\chi_{mol} = \chi_0 + C/(T-\theta)$, where $\chi_0$ is the temperature independent part of susceptibility, $C$ is Curie constant. From the Curie constant we have obtained the effective moments by $\mu_{eff} = \sqrt{\frac{3k_BC}{N_A}}$ where $N_A$ is the Avogadro's number and $k_B$ is the Boltzmann constant. The obtained $\mu_{eff}$ ranges from 0.67 to 1.05 $\mu_B$ (Fig. 2e), close to the theoretically expected value of 0.86 $\mu_B$ for a $Sm^{3+}$ ion with a $4f^5$ configuration. In addition to the uncertainty from the fitting, the deviation from the expected value could be due to a few reasons. For example, the polarization of conduction electrons or the reduction of moment density in frustrated magnetic systems can lead to an overestimates, while the localized 4*f* electrons screened by the conduction electrons or the reduction of spin-orbit coupling strength due to structure distortion can cause moment reduction [46–49]. From the fitting we also extracted negative $\theta_{cw}$ which is expected for AFM ordering. The fitted $\theta_{cw}$ exhibits a non-monotonic dependence on Sb content *x*, becoming maximal (more negative) when *x* approaches the two ending compositions (i.e., $x = 0$ and 1). For these ending compositions, because $T_N$ also reduces (Fig. 2d), the frustration parameter, defined as $f = |\theta_{CW}|/T_N$, is significantly enhanced (around 7 ~ 8), which is suggestive of possible magnetic frustration in $SmSb_xTe_{2-x}$ compounds. This may originate from a competition between the nearest-neighbor and next-nearest-neighbor magnetic exchange coupling in those compounds as suggested earlier [50–52].

Figures 3a-c present the isothermal magnetization *M*(*H*) measurements on $SmSb_xTe_{2-x}$. The stoichiometric SmSbTe has been reported to show linear *H*-dependence for *M* with in-plane field and slight non-linearity when *H*//*c* [36]. In non-stoichiometric $SmSb_xTe_{2-x}$ studied in this work, the non-linear *M*(*H*) becomes more obvious upon reducing the Sb content for both magnetic field orientations. The isothermal magnetization measured under both in-plane and out-of-plane field orientations for an orthorhombic Sb-less $SmSb_{0.13}Te_{1.91}$ sample and a tetragonal Sb-rich $SmSb_{0.93}Te_{1.10}$ sample are shown in Fig. 3c for comparison, from which clear non-linear *M*(*H*) similar to the reported metamagnetic transitions in CeSbTe [27,28,30,33] and GdSbTe [28,35] can be seen, implying similar metamagnetic transitions in $SmSb_xTe_{2-x}$ compounds. The linear *M*(*H*) in *H*//*ab* measurement and a small deviation from linearity along *H*//*c* measurements is observed for Sb rich compounds. The deviation from linearity is increasing with lowering the Sb content, which implies the correlations between magnetism and structure in SmSbTe system. Such a correlation

between distorted Sb-square net and magnetization has been observed in off-stoichiometric compound $GdSb_xTe_{2-x-\delta}$ [29,35].

Magnetism in $SmSb_xTe_{2-x}$ has also been studied by heat capacity measurements. As shown in Fig. 4a, heat capacity reveals magnetic phase transitions consistent with those probed in magnetic susceptibility measurements (Fig. 2a). Similar to the stoichiometric SmSbTe which shows a broad hump at the magnetic transition temperature in specific heat divided by temperature data $C(T)/T$ [36], the tetragonal non-stoichiometric $SmSb_xTe_{2-x}$ ($x > 0.8$) samples also display a broad specific peak. As mentioned above, with reducing the Sb content below the tetragonal-orthorhombic phase boundary, the magnetic susceptibility measurements have revealed multiple magnetic ordering temperatures for $0.36 \leqslant x \leqslant 0.8$ (Figs. 2a-b). The feature can be clearly resolved in $C(T)/T$. As shown in Fig. 4a, an additional sharp peak appears at higher temperatures (6 ~ 7 K) when $x$ drops below 0.8. Such sharp peak is the strongest near $x = 0.8$, which gradually shifts to lower temperatures and become suppressed with further reducing the Sb content. The temperatures for the broad ($T_{N1}$) and sharp ($T_{N2}$) heat capacity peaks agree well with the observations in magnetic susceptibility measurements, as summarized in Fig. 2d. However, the sharp peak in $x = 0.13$ sample might have a different mechanism compared to the sharp $T_{N2}$ peak of $x = 0.36 – 0.78$ samples. As shown in Fig. 4a, the tetragonal samples ($x > 0.8$, bottom of Fig. 4a) show only one broad peak ($T_{N1}$). With decreasing the Sb content to the tetragonal-orthorhombic phase boundary ($x \sim 0.8$), in addition to the broad $T_{N1}$ peak, a sharp and strong $T_{N2}$ peak appears. With further lowering the Sb content, the broad $T_{N1}$ peak can be observed in all samples except for the $x = 0.13$ sample, which is the sample with the lowest Sb content in this work. For the $T_{N2}$ peak, it gradually shifts to lower temperature, becomes broadened, and suppressed. The $T_{N2}$ peak finally disappears for the $x = 0.25$ sample. The broadening of the peak and the suppression of the peak amplitude is systematic. For the $x = 0.13$ sample, a sharp peak is observed. However, the sharpness and large amplitude does not follow the trend of the composition dependence of the $T_{N2}$ peak. Therefore, we conclude that such sharp peak may have a mechanism different from the $T_{N2}$ peak in other samples. For example, a different magnetic transition or coupling with a possible structure transition or ordering of sub-lattice or vacancies. Though the broad specific heat peaks in various $SmSb_xTe_{2-x}$ coincide well with the magnetic transition temperatures probed in susceptibility measurements, the absence of well-defined peaks implies possible magnetic frustration in this materials system [50,53,54]. A broad hump below the ordering temperature owing to the partial disorder of the spins due to structural defects has been reported in heavy fermion systems [55–60]. Furthermore, a broad peak arising from Schottky anomaly due to the crystal field splitting or the ordering of the rare earth ions has also been observed in several rare-earth compounds [61,62]. To clarify the mechanism for the broad specific peak in $SmSb_xTe_{2-x}$, more theoretical and experimental efforts are needed.

The measurements under different magnetic fields reveal field-independent heat capacity for $SmSb_xTe_{2-x}$ with various $x$. An example for Sb-less, orthorhombic $SmSb_{0.72}Te_{1.30}$ is shown in Fig. 4b, in which the peak positions and amplitudes for both specific heat peaks do not vary with field. For Sb rich ($x > 0.8$) tetragonal samples, the same behavior has also been observed in our previous work on stoichiometric SmSbTe [36], which shows only one broad, field independent heat capacity peak.

To extract electronic ($C_e$) and magnetic ($C_m$) specific heat, we used the nonmagnetic isomorphous compound LaSbTe as a reference sample to evaluate the phonon contribution ($C_{ph}$). We adopted the principle of corresponding states which has been used for specific heat analysis for stoichiometric SmSbTe [36]. The effectiveness of this approach has also been demonstrated by heat capacity studies on other compounds such as NdSbTe and Fe(Te, Se) [31,63]. An example of fitting the electronic ($C_e/T$) and phonon ($C_{ph}/T$) specific heat at temperatures above the magnetic ordering temperatures for $SmSb_{0.72}Te_{1.30}$ is shown by the red line in Fig. 4b, from which the Sommerfeld coefficient $\gamma$ and Debye temperature $\theta_D$ can be obtained from the low temperature extrapolation of the fit [36]. Fig. 4c summarizes the evolution of $\gamma$ and $\theta_D$ with the Sb content $x$ in $SmSb_xTe_{2-x}$. Relatively large $\gamma$ (80 – 165 mJ/mol K$^2$) has been obtained, which displays a non-monotonic composition dependence with a minimum value near $x = 0.4$ and maximum values near the two ending compositions ($x = 0$ and 1). The large $\gamma$ values for $SmSb_xTe_{2-x}$ are consistent with that for stoichiometric SmSbTe [36], which has also been observed in some *Ln*SbTe materials such as NdSbTe (115 mJ/mol K$^2$) [31] and HoSbTe (383.2 mJ/mol K$^2$) [30]. One possible mechanism is the presence of the flat Sm 4*f* bands near the Fermi level which may hybridize with the conduction electrons and lead to mass enhancement [36]. It is worth noting that large $\gamma$ is not a generic feature in *Ln*SbTe compounds. The magnetic CeSbTe [27,33] and GdSbTe [28], as well as the non-magnetic LaSbTe [31] all display small Sommerfeld coefficients. Compared to the large variation for $\gamma$ with $x$ in $SmSb_xTe_{2-x}$, Debye temperature only exhibits a small increase from 220 K to 250K upon increasing the Sb content (Fig. 4c). Such weak composition dependence for $\theta_D$ implies small structure modifications with Sb substitution for Te in $SmSb_xTe_{2-x}$, which is consistent with our structure analysis stated above (Fig. 1d).

The magnetic specific heat $C_m/T$ for various $SmSb_xTe_{2-x}$, obtained from subtracting the fitted electronic and phonon contributions from the total measured specific heat, is shown in Fig. 4d. The corresponding magnetic entropy can be evaluated by $S_m = \int_0^T \frac{C_{m}(T)}{T} dT$. To calculate $S_m$, we took $C_m/T = 0$ at zero temperature. As shown in Fig. 4e, magnetic entropy for various $SmSb_xTe_{2-x}$ increases quickly with increasing temperature and saturates to 3.9 – 4.8 J/mol K above $T$ = 10 K. Such values are less than the expected magnetic entropy of $R\ln2$ = 5.76 J/mol K ($R$ is the molar gas constant) for a $J$ = 5/2 doublet for $Sm^{3+}$, which implies possible residual magnetic entropy that has been widely seen in frustrated systems [64,65]. Indeed, in the crystals we have investigated, the lowest $S_m$ (i.e., the entropy missing is the greatest) is observed in $SmSb_{0.78}Te_{1.20}$, which is very close to the orthorhombic-tetragonal phase boundary. Such a boundary sample might possess strong frustrations as the result of competition between two structural phases.

To obtain a complete understanding of the electronic and magnetic properties for $SmSb_xTe_{2-x}$, we have characterized the electronic transport properties. Figure 5 show the temperature dependence of in-plane resistivity $\rho_{xx}(T)$ for a few compositions, which is normalized to the resistivity at $T$ = 300 K for better comparison. Overall, a non-metallic transport behavior is observed in all compositions, similar to many *Ln*SbTe (*Ln* = Ce, Gd, Nd) materials [27,28,31,34]. The low Sb content samples $SmSb_{0.24}Te_{1.61}$ and $SmSb_{0.34}Te_{1.88}$ displayed an insulating-like behavior with an abrupt resistivity upturn at low temperatures. Such upturn disappears with increasing the Sb content in $SmSb_xTe_{2-x}$, implying an enhancement of metallicity though the overall temperature dependence for resistivity still shows a non-metallic behavior. The enhanced metallicity with increasing Sb content has also been observed in $GdSb_xTe_{2-x}$ [26] . Furthermore, similar to $GdSb_xTe_{2-x}$ [26], resistivity for $SmSb_xTe_{2-x}$ does not display any clear feature at $T_N$, implying the spin scattering might not play a very important role for electron transport in these materials.

## 4. Conclusion

In summary, we have studied the composition dependence of structural, magnetic, thermal dynamical, and electronic transport properties for $SmSb_xTe_{2-x}$. The coincidence of tetragonal-to-orthorhombic structure transition, emergence of multiple magnetic transitions, and magnetic entropy reduction implies the coupling between structure and magnetism in this material system. Given the existence of Dirac states in stoichiometric SmSbTe and the robustness of the topological states against orthorhombic distortions in other related compounds in this material family [26,34], the demonstrated tuning of metallicity with Sb content and the enhanced electron effective mass in this work further suggest $SmSb_xTe_{2-x}$ as a good platform for engineering quantum states.

## Acknowledgments

Work at the University of Arkansas (synthesis and characterizations) is supported by U.S. Department of Energy Office of Science under Award No. DE-SC0022006. R.B. acknowledges the support from the Chancellor's Innovation and Collaboration Fund at the University of Arkansas (for personnel support). B.D. thanks the support by JSPS KAKENHI Grant Number JP21K14656 and by Grants for Basic Science Research Projects from The Sumitomo Foundation.

**Table 1** Crystallographic data for various orthorhombic (orth.) and tetragonal (tet.) $SmSb_xTe_{2-x}$. obtained from the refinement of single crystal XRD. The data were collected at room temperature using graphite-monochromated Mo-$K_\alpha$ radiation ($\lambda$ =0.71073 Å).

| | **Space group** | **Lattice constants (Å)** | | | **Atomic positions** | | | | | | **Goodness of fit** |
|---|---|---|---|---|---|---|---|---|---|---|---|
| | | ***a*** | ***b*** | ***c*** | **Atom** | **Wyckoff** | ***x*** | ***y*** | ***z*** | ***U*<sub></sub>**$_{\mathbf{eq}}$ | |
| $SmSb_{0.11}Te_{1.85}$ | *Pmmn* Orth. | 4.3146 (6) | 4.3847 (7) | 9.0582 (10) | Sm | 2a | 0.2500 | 0.2500 | 0.7726 | 0.016 | 1.277 |
| | | | | | Te | 2a | 0.2500 | 0.2500 | 0.1298 | 0.015 | |
| | | | | | Sb | 2b | 0.2500 | 0.7500 | 0.5025 | 0.028 | |
| $SmSb_{0.38}Te_{1.51}$ | *Pmmn* Orth. | 4.3236 (6) | 4.3838 (7) | 9.0629 (17) | Sm | 2b | 0.2500 | 0.7500 | 0.2721 | 0.022 | 1.237 |
| | | | | | Te | 2b | 0.2500 | 0.7500 | 0.6294 | 0.021 | |
| | | | | | Sb | 2a | 0.2500 | 0.2500 | 0.0026 | 0.033 | |
| $SmSb_{0.62}Te_{1.37}$ | *Pmmn* Orth. | 4.3008 (3) | 4.3432 (3) | 9.1570 (6) | Sm | 2a | 0.2500 | 0.2500 | 0.2744 | 0.016 | 1.283 |
| | | | | | Te | 2a | 0.2500 | 0.2500 | 0.6280 | 0.015 | |
| | | | | | Sb | 2b | 0.2500 | 0.7500 | 0.0011 | 0.028 | |
| $SmSb_{0.83}Te_{1.04}$ | *P*4/*nmm* Tet. | 4.2994 (2) | 4.2994 (2) | 9.2521 (6) | Sm | 2c | 0.2500 | 0.2500 | 0.7759 | 0.004 | 1.327 |
| | | | | | Te | 2c | 0.2500 | 0.2500 | 0.1264 | 0.004 | |
| | | | | | Sb | 2b | 0.7500 | 0.2500 | 0.5000 | 0.006 | |
| $SmSb_{0.93}Te_{1.07}$ | *P*4/*nmm* Tet. | 4.2978 (2) | 4.2978 (2) | 9.2630 (6) | Sm | 2c | 0.2500 | 0.2500 | 0.7244 | 0.016 | 1.550 |
| | | | | | Te | 2c | 0.2500 | 0.2500 | 0.3734 | 0.019 | |
| | | | | | Sb | 2a | 0.7500 | 0.2500 | 1.0000 | 0.017 | |

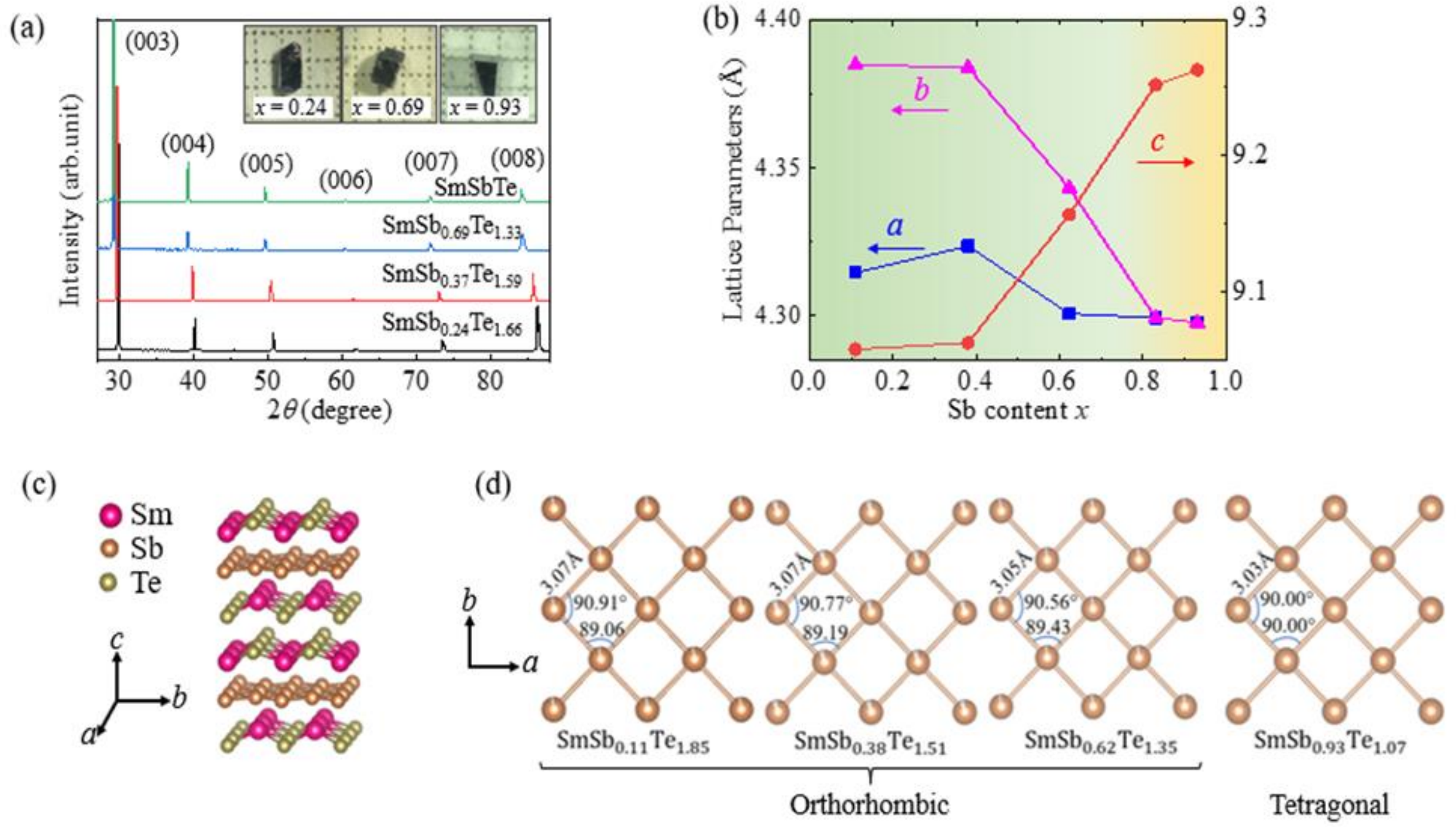

**Fig. 1** (a) X-ray diffraction patterns for $SmSb_xTe_{2-x}$ single crystals, showing the (00*l*) reflections. Inset: images of the single crystals with various Sb content. The square mesh measures 1 $mm^2$. (b) Evolution of lattice parameters with varying Sb content. The green and yellow regions represent orthorhombic and tetragonal lattices, respectively. (c) Crystal structures for $SmSb_xTe_{2-x}$. The structure parameters are provided in Table 1. Note that the tetragonal ($x > 0.8$) and orthorhombic ($x > 0.8$) structures looks very similar so only one structure is shown. The difference of two structures is presented in (d). (d) Distorted Sb planes in orthorhombic $SmSb_{0.11}Te_{1.85}$, $SmSb_{0.38}Te_{1.51}$, $SmSb_{0.62}Te_{1.35}$, and the undistorted Sb square-net in tetragonal $SmSb_{0.93}Te_{1.07}$.

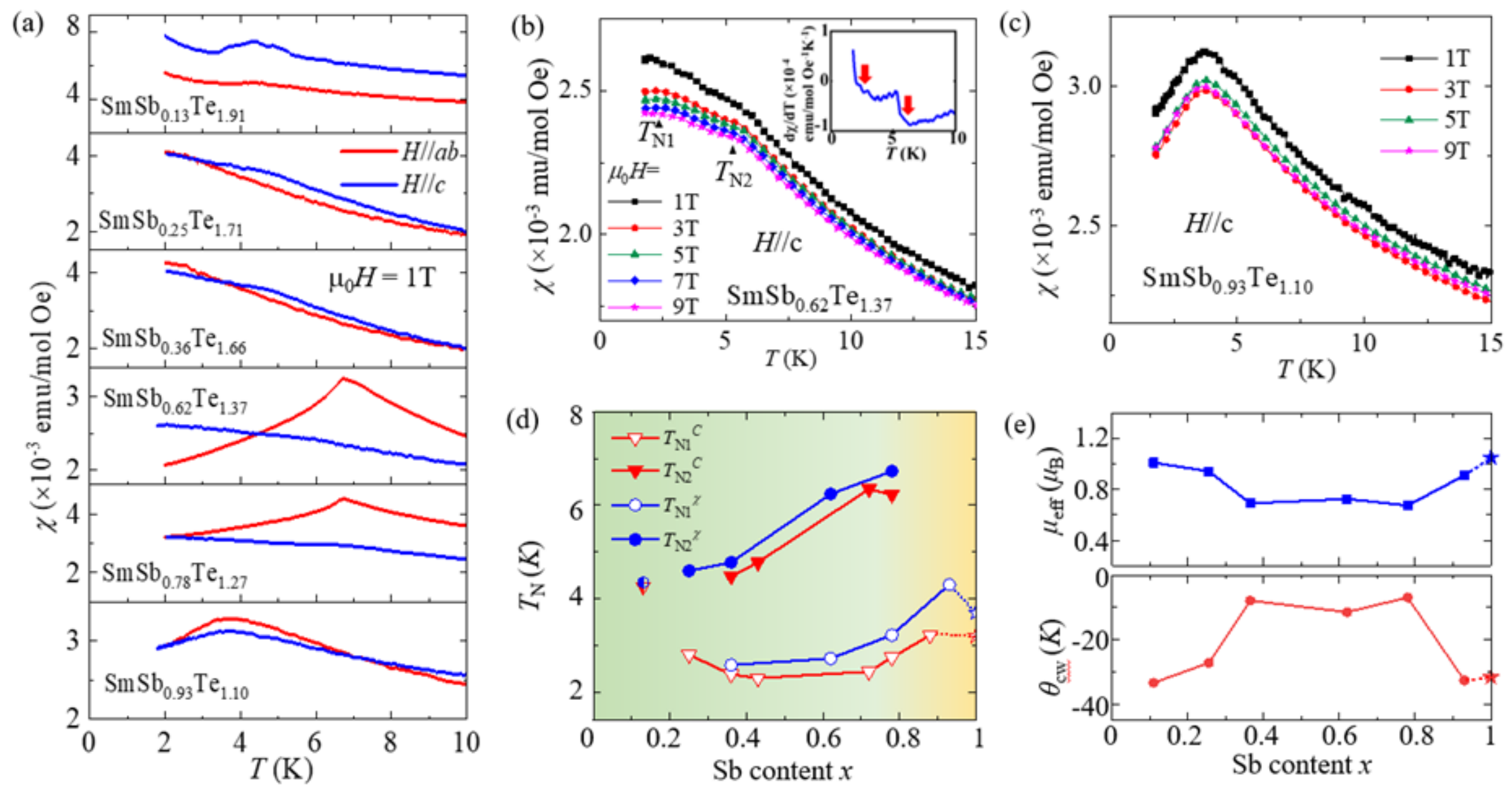

**Fig. 2**. (a) Temperature dependent molar susceptibility $\chi$ for $SmSb_xTe_{2-x}$ measured under in-plane (*H*//*ab*) and out-of-plane (*H*//*c*) magnetic field of 1T. (b) Temperature dependent molar susceptibility of $SmSb_{0.62}Te_{1.37}$ measured at different out-of-plane magnetic fields from 1 to 9 T. Inset: the derivative of the 9T susceptibility $d\chi/dT$. The two successive magnetic transitions are indicated by the red arrows. (c) Temperature dependent molar susceptibility for $SmSb_{0.93}Te_{1.10}$ measured at different out-of-plane magnetic fields from 1 T to 9 T. (d) Evolution of the magnetic transition temperatures $T_{N1}$ and $T_{N2}$ with varying Sb content, extracted from magnetization (labeled as $T_N^{\chi}$) and heat capacity (labeled as $T_N^{C}$) measurements. The green and yellow regions represent orthorhombic and tetragonal lattice symmetries, respectively. (e) Evolution of effective magnetic moment $\mu_{eff}$ and Curie-Weiss temperature $\theta_{CW}$ with varying Sb content. Data for the $x = 1$ sample in (d) and (e) is taken from Ref. [36] for comparison.

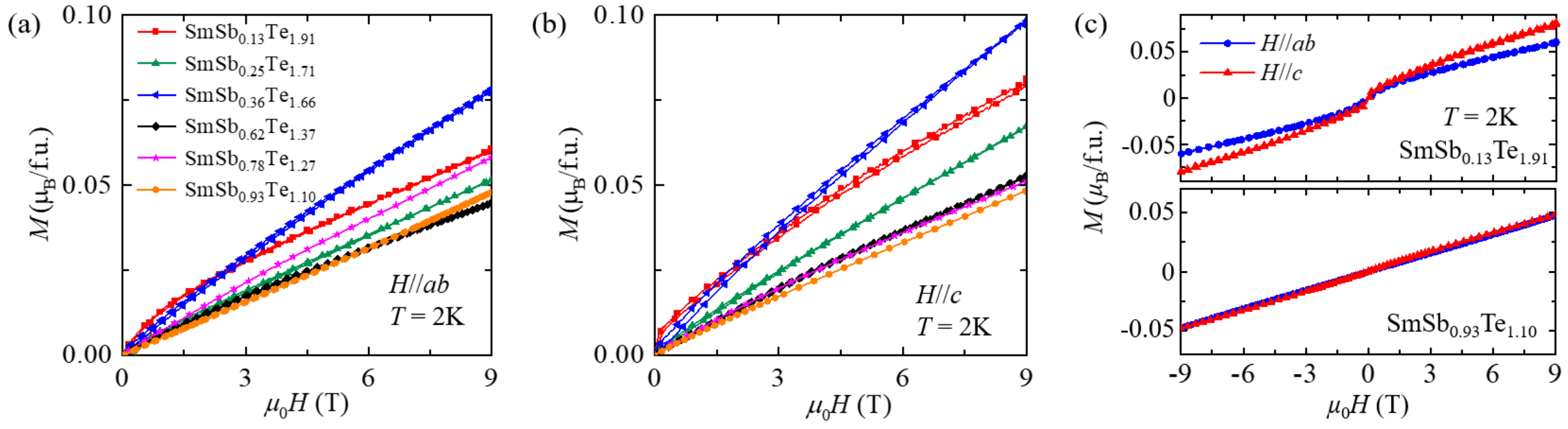

**Fig. 3** (a-b) Field dependent isothermal magnetization of various $SmSb_xTe_{2-x}$ at $T$ = 2 K measured under (a) in-plane ($H//ab$) and (b) out-of-plane ($H//c$) magnetic field orientations. Same color codes are used in (a) and (b). (c) Comparison of isothermal magnetization measurements for $SmSb_{0.13}Te_{1.91}$ and $SmSb_{0.93}Te_{1.10}$ at $T$ = 2 K under in-plane and out-of-plane magnetic fields.

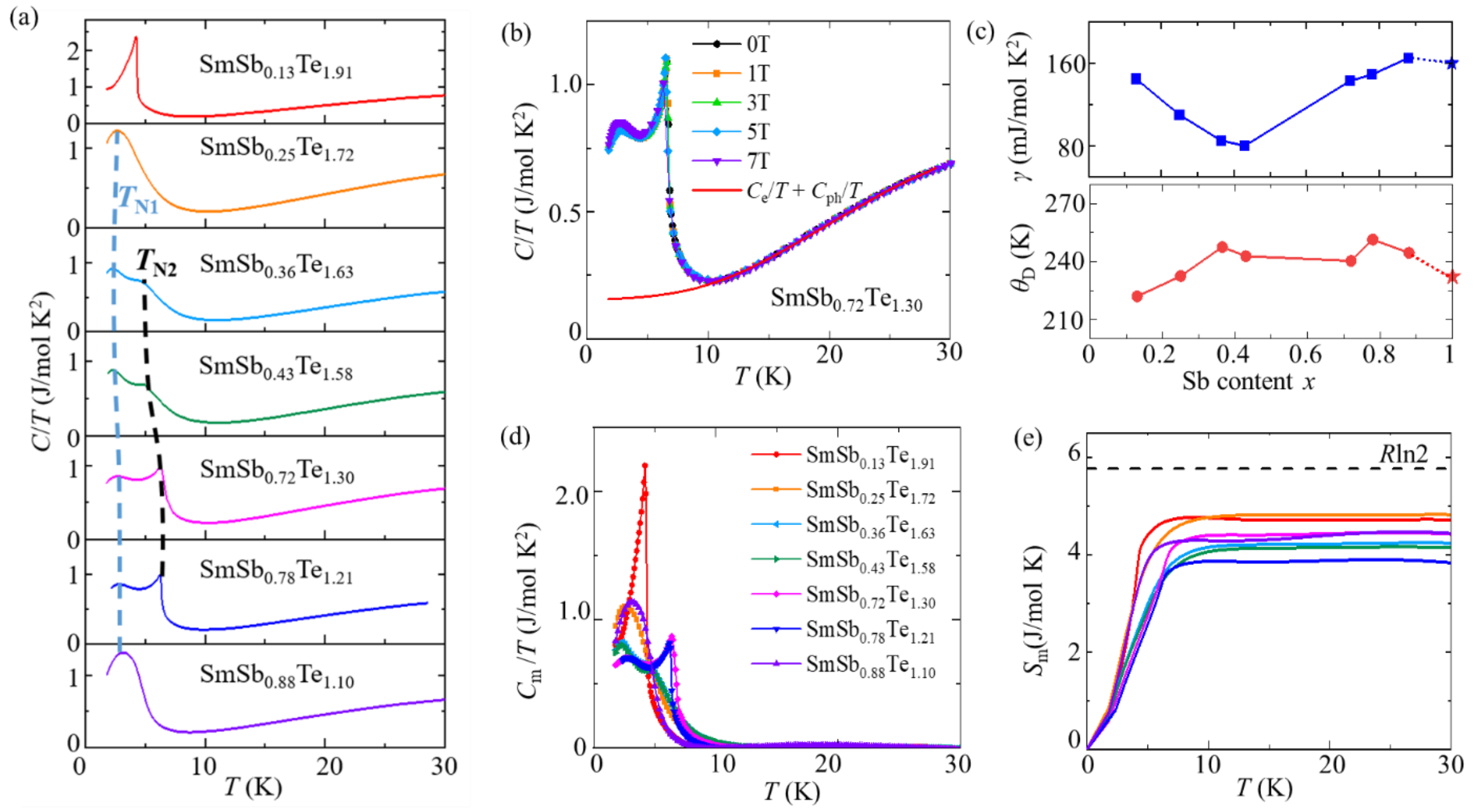

**Fig. 4** (a) Temperature dependent specific heat divided by temperature $C/T$ for $SmSb_xTe_{2-x}$. The blue and black dotted line indicate the evolution of $T_{N1}$ and $T_{N2}$ peaks with Sb content. (b) Specific heat divided by temperature for $SmSb_{0.72}Te_{1.30}$ measured under various magnetic fields from 0 to 7 T. The red line indicates the electronic and phonon contributions to specific heat, obtained from fitting the data using LaSbTe as a reference sample (see text). (c) Evolution of Sommerfeld coefficient $\gamma$ and Debye temperature $\theta_D$ with varying Sb content $x$. Data for $x = 1$ sample in (d) and (e) is taken from Ref. [36] for comparison. (d) Magnetic specific heat for various $SmSb_xTe_{2-x}$. (e) Magnetic entropy $S_m$ of $SmSb_xTe_{2-x}$. Same color codes are used in (a), (d), and (e).

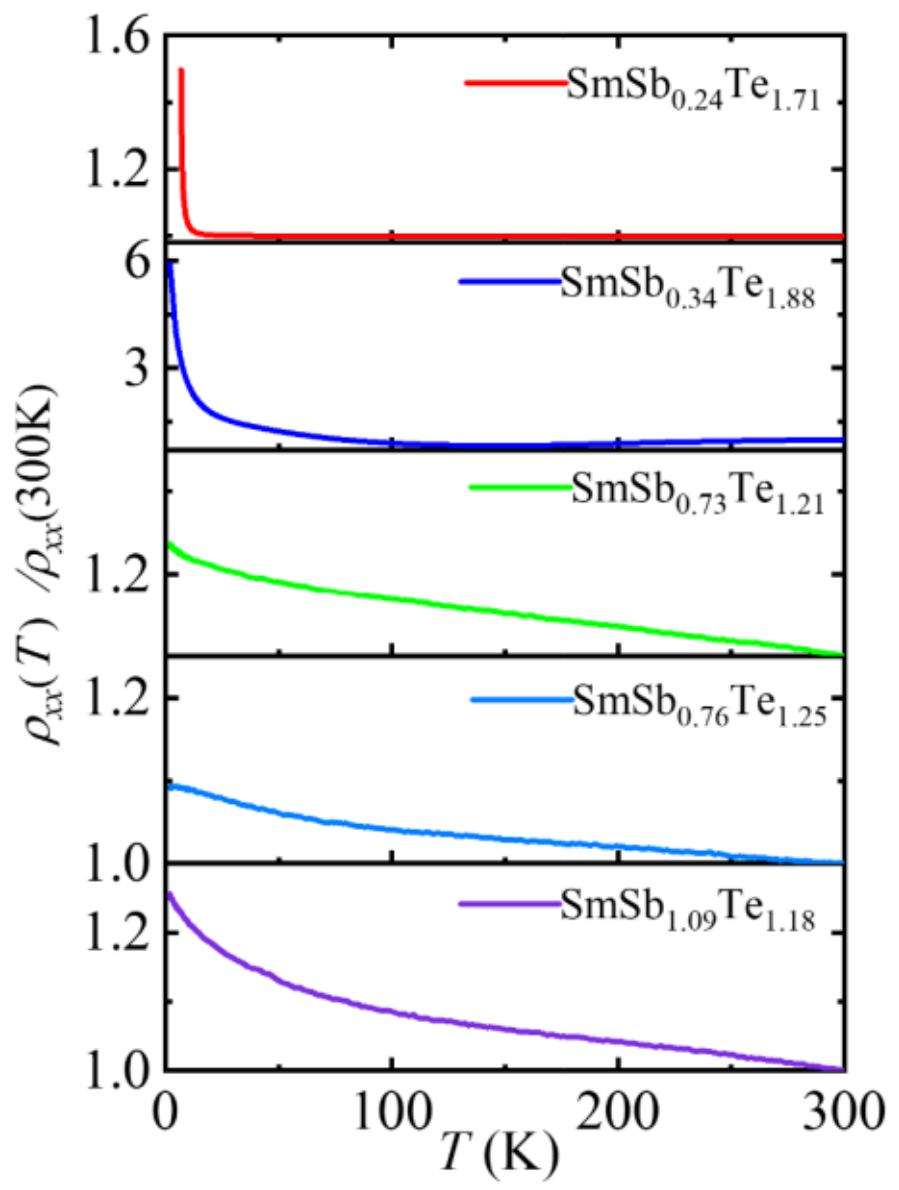

**Fig. 5** Temperature dependent in-plane resistivity for various $SmSb_xTe_{2-x}$.